\documentclass[journal]{IEEEtran}
\pdfoutput=1

\ifCLASSINFOpdf
  \usepackage[pdftex]{graphicx}
  \graphicspath{{./}}
  \DeclareGraphicsExtensions{.pdf}
\else
\fi

\usepackage{dsfont}
\usepackage[cmex10]{amsmath}

\usepackage{booktabs}
\usepackage{array}

\begin{document}

\title{Modeling Weather Conditions Consequences\\ on Road Trafficking Behaviors}

\author{Guillaume~Allain,~\IEEEmembership{Mediamobile,}
        Fabrice~Gamboa,~\IEEEmembership{Institut de Math\'ematiques de Toulouse,}
        Philippe~Goudal,~\IEEEmembership{Mediamobile,}
        Jean-No\"el~Kien,~\IEEEmembership{Institut de Math\'ematiques de Toulouse \& Mediamobile,}
        and~Jean-Michel~Loubes,~\IEEEmembership{Institut de Math\'ematiques de Toulouse}
\thanks{This work was supported by \mbox{Mediamobile} with a partnership with \mbox{M\'et\'eo-France}.}
\thanks{Jean-No\"el~Kien is with the Research and Innovation Division at \mbox{Mediamobile}, 27 boulevard Hippolyte Marqu\`es, 94200 Ivry-sur-Seine, France e-mail: \mbox{Jean-Noel.KIEN@v-trafic.com}}
\thanks{Guillaume~Allain is Director of Contents Division at \mbox{Mediamobile}}
\thanks{Fabrice~Gamboa is Professor at Institut de Math\'ematiques de Toulouse}
\thanks{Philippe~Goudal is Director of Research and Innovation Division}
\thanks{Jean-Michel~Loubes is Professor at Institut de Math\'ematiques de Toulouse}
}


\maketitle

\begin{abstract}
We provide a model to understand how adverse weather conditions modify traffic flow dynamic.
We first prove that the microscopic Free Flow Speed of the vehicles is changed and then provide a rule to model this change.
For this, we consider a thresholded linear model, corresponding to an application of a MARS model \cite{Friedman1993} to road trafficking.
This model adapts itself locally to the whole road network and provides accurate unbiased forecasted speed using live or short term forecasted weather data information.
\end{abstract}

\begin{IEEEkeywords}
Forecasting method, Linear thresholded model, Road traffic, Spatial extrapolation, Weather.
\end{IEEEkeywords}

\IEEEpeerreviewmaketitle

\section{Introduction}

\IEEEPARstart{I}{t} is commonly accepted that adverse weather conditions modify significantly traffic flow dynamics in a complex way. Actually, it is well known that bad weather conditions such as, heavy rain, fog, snow, induce a significant decrease on  traffic flow speeds. Note that it can be partially explained by the legal speed regulations. However, if several studies conclude that road traffic speed decreases during adverse weather, this trend is only confirmed. Furthermore, up to our knowledge, no quantitative analysis has been conducted to forecast the evolution of the observed speed of vehicles. Even deterministic models for road traffic fail in this study since they usually involve a large set of parameters which are all affected by the change of weather conditions. This prevents the use of such equations. In this paper, we tackle this issue and provide a general model to estimate the change in traffic flow speed for different weather scenarios.\\

Actually, trying to understand the impact of weather conditions requires a direct comparison between observed traffic speeds whose variability is only due to such changes.  Hence, quantifying the impact of adverse weather conditions on traffic speed can only be conducted through the analysis of a two paired data. Each pair corresponds to {\it similar} traffic conditions but with different weather conditions.  This is a difficult issue since road trafficking is a non stationary phenomenon and much of  the variability is due to the road condition changes (with the occurrence of traffic jams). So, using empirical studies is not obvious because of the heterogeneity of the data. Moreover, some conditions are scarcely observed. This increases the difficulty of the estimation and requires to choose a set of weather conditions that makes sense for the road manager but also with a large enough number of observations.\\
Some work has already been conducted in this direction. We refer to \cite{ElFaouzi2008},\cite{Elfaouzi2010},\cite{Elfaouzi2010Bis},\cite{Billot2009},\cite{Billot2010} and \cite{Kilpelainen2007}. Nevertheless, the drawback of previous methods is that the speed modifications are global. This means that the speeds are affected without enabling these changes to depend on the values of the initial velocity. In this work, we use a more flexible model using Multivariate Adaptive Regression Splines (MARS)  model to consider thid initial condition. Furthermore, using a threshold enables to consider different impacts according to different levels  of weather changes. Such procedure has been described in \cite{Friedman1991} and its implementation is detailed in \cite{Friedman1993}. Some work was conducted in the same direction, see for instance \cite{Garthwaite1997}, \cite{Mammen1997} and references therein. The calibration of the model is achieved by estimating the parameters on a learning set. This last set is built by selecting pairs of observations under the same traffic condition but with a different weather condition.  We study the performance of this model and prove that it enables to forecast accurately the possible speed evolutions.\\

The paper falls into the following parts. Section~\ref{s:one} is devoted to the description of the data and their particularity. The following section, Section~\ref{s:two} describes the construction of the model while its performances are analyzed in Section~\ref{s:three}. Finally, we discuss some of our results and draw some guidelines for speed forecast under adverse weather conditions in Section~\ref{s:conclusion}.

\section{Data and Issues}\label{s:one}

\subsection{Description}
\label{ss:description} 
A road network can be represented as a set of links connected together in a form depending on the underlying road network. Usually, links are classified by a well-known attribute: the Functional Road Class (FRC) \cite{Poovadan2011}. FRC is a classification based on the importance of the road in the connectivity of the total network. Table \ref{FullNameandAttributeValues} recaps the relation between the FRC and the network. In our study, industrial constraints make us work only on roads from FRC 0 to 2 because the most part of the information provided by Mediamobile that matches the customer demand concerns this FRC range.\\

\begin{table}[!ht]
  \caption{Full name and attribute values by FRC values}
  \label{FullNameandAttributeValues}
  \centering
  \begin{tabular}{c l}
    \toprule %
    FRC & Full name and Attribute values\\
    \toprule %
    0 & Motorway, Freeway, or other Major road\\
    1 & Major road less important than a motorway\\
    2 & Other major road\\
    3 & Secondary road\\
    4 & Local connecting road\\
    5 & Local road of high importance\\
    6 & Local road\\
    7 & Local road of minor importance\\
    8 & Other road\\\bottomrule
  \end{tabular}
\end{table}

Vehicles equipped with some GPS device can return at each time their positions to a server. So that, they may be considered as floating sensors on the network. Such sensors form the Floating Car Data (FCD) source of traffic information. A map-matching algorithm (out of the scope of this article) establishes speed $\mathcal{V}(x,t)$ on a link $x$ at a time $t$ from a couple of successive positions and times by matching them on a digitized network. Then, when a significant number of speeds on a link $x$ are at hand, we may produce a microscopic processed FCD speed $\mathcal{V}(x,t)$. By using this gathering technology, we are not geographically dependent of any counting station but we are limited by the GPS users feedback. Nevertheless, this source of traffic data is relevant since we have a huge amount of data (\mbox{515 798 606} positions in March 2011 for instance) and we potentially cover all the network. The traffic database used in the paper was provided by Mediamobile and is composed of the vehicle speeds over $T = 107$ days.\\

The weather database used in this paper was also provided by Mediamobile through his partnership with M\'et\'eo-France. Since 2009, M\'et\'eo-France provides to Mediamobile a new high quality flow of data. It incorporates geo-tagged point every unit of 120 000 road kilometers of the France network. This flow of meteorological data is an aggregation of real time and forecasted observations. Then for our study, we have at hand a regular flow of weather data $\mathcal{M}(x,t)$ on a link $x$ at a time $t$ updated every 15 minutes. In this paper, we will focus on the following bad weather conditions: soft rain, medium and strong rain, rain and snow mixed, drizzle.\\

\subsection{Data quality}

Individual microscopic traffic data usually exhibit high noise and outliers due to several causes

\begin{itemize}
\item GPS logger accuracy,
\item incorrect vehicle path estimation: a wrong projection of the vehicle path can lead to incorrect speeds and further, speeds on incorrect links,
\item inner variations of individual vehicles in the traffic flow.
\end{itemize}

To decrease the noise and eliminate outliers, we use a two step filtering algorithm:

\begin{enumerate}
\item the first step filters out  aberrant data. Mediamobile estimates a Free Flow Speed (FFS) defined as the most likely speed in free flow traffic conditions. Then, we filter out aberrant data where speed records are higher than 150\% of the reference speed i.e. the FFS,
\item the second step put out links that do not have enough records. Here, we arbitrarily fix to 100 measures the minimum amount of data necessary to keep a link.
\end{enumerate}

After performing this algorithm, we have confident traffic data. Weather data are already consistent because they have been preprocessed by M\'et\'eo-France. So that, they do not need any more treatment.

\subsection{Issues}

The main issues are twofold:
\begin{itemize}
 \item building a learning set for weather condition. Actually, the main goal is to build couples of speeds at a given location observed under the same traffic condition but with a different weather condition in order to understand the weather conditions consequences on road trafficking behaviors. First, we need to associate both traffic data and weather data. The frequency of weather data flow is 15 minutes. So even if a weather condition is observed at a time $t_0$ only, we will propagate it to the whole interval $[t_0, t_0 + 15[$ such that all speeds $\mathcal{V}(x,t)$ observed in this interval get paired with $\mathcal{M}(x,t_0)$,

 \item finding a predictive rule to forecast velocities. We used the heuristic idea that adverse weather conditions do not affect the velocity in the same way. Indeed, we build a model that includes a different treatment for different ranges of speeds. This rule must be stable to be extended to the whole road network, yet providing good enough estimations.
\end{itemize}

\section{Modeling } \label{s:two}
\subsection{Regression}
\label{regression}
\indent Our aim is to rectify the forecasted speed according to road weather conditions. Many methods exist to forecast speeds on a road network but this feature is out of the scope of this article. Some examples of such methods can be find in \cite{Loubes2006} and \cite{Allain2009}. In this work, we already have forecasted speed at hand and we want to correct them according to road weather conditions. This will be done by applying a correction on the forecasted speed. We focus on a bias depending on the speed. Indeed, the expertise of road trafficking theory shows that under adverse road weather conditions, drivers reduce their velocities when they go fast whereas they do not change it otherwise.\\
An usual way to rectify the speed is obtained by using a polynomial function of degree 1

\begin{equation}
\label{MARS}
\begin{array}{l}
\displaystyle \forall t\in \mathcal{V}(t_0),\hspace{2em}  V(x,t) =\\
\displaystyle \sum^p_{k=1} P_{\theta_k}\Big( V(x,t_0) \Big)\times\mathds{1}_{[\alpha_k,\alpha_{k+1}]}\Big(V(x,t_0) \Big)
\end{array}
\tag{$\mathcal{M}_{1}$}
\end{equation}

with $\theta_k = [a_k;b_k]$, $P_{\theta_k}$ a polynomial of degree 1 such that $P_{\theta_k}\Big(V(x,t_0)\Big) = a_k . V(x,t_0) + b_k$ and level of speeds $\forall k \in [1..p],\hspace{2em} [\alpha_k;\alpha_{k+1}]$ are such that $\alpha_1 = 0$ and $\alpha_{p+1} = +\infty$. $V(x,t_0)$ is the forecasted speed at time $t_0$ and $V(x,t)$ its corrected speed.  $\mathcal{V}(t_0)$ is a short term neighborhood around $t_0$. This means that we will adjust $V(x,t_0)$ in a time neighborhood while no newer speed data is available.\\

This model is no more than the so-called Multivariate Adaptive Regression Splines (MARS) introduced in ~\cite{Friedman1993} where non linearities match driver's behavior changes. We refer to this model \ref{MARS} as the MARS model. Indeed, we can rewrite the model \ref{MARS} in a classical MARS form as

\begin{equation*}
  \displaystyle \forall t\in \mathcal{V}(t_0),\hspace{2em}  V(x,t) = \sum^p_{k=1} Q_{\Gamma_k}\Big( h_{1k}, h_{2k}, h_{3k} \Big)
\end{equation*}

with $\Gamma_k = [\gamma_{1k};\gamma_{2k};\gamma_{3k}]$, $Q_{\Gamma_k}$ a polynomial of degree 1 such that
$Q_{\Gamma_k}\Big( h_{1k}, h_{2k}, h_{3k} \Big) = \gamma_0 + \gamma_{1k}.h_{1k}+ \gamma_{1k}.h_{2k}+ \gamma_{1k}.h_{3k}$
, a set of hinge functions
\[
\left \{
  \begin{array}{c c l}
    h_{1k} & = & \max(0, V(x,t_0) - \alpha_k) \\
    h_{2k} & = & \max(0, V(x,t_0) - \alpha_{k+1}) \\
    h_{3k} & = & \max(0, \alpha_{k+1} - V(x,t_0))
  \end{array}
\right.
\]
and level of speeds $\forall k \in [1..p], \hspace{2em} [\alpha_k;\alpha_{k+1}]$ such that $\alpha_1 = 0$ and $\alpha_{p+1} = +\infty$.\\

We point out in Section \ref{stability} that the model \ref{MARS} suffers from a lack of stability on the network. Indeed, it has not an homogeneous structure among the network. This means that we are not able to extrapolate the model to all links. Moreover this model may apply a correction on low speeds under adverse weather condition. Hence, it does not fit the important feature: driving at low
speed under bad weather conditions has no impact on drivers' behavior. That is why, we rather use the following variant of model \ref{MARS} that fits better the traffic flow theory and that is easier to extrapolate

\begin{equation}
\label{ltm}
\begin{array}{l}
\displaystyle \forall t\in \mathcal{V}(t_0),\hspace{2em}  V(x,t)  = \\ 
\displaystyle V(x,t_0) \times \mathds{1}_{[0,\frac{\theta_0}{1-\theta_1}[}\Big(V(x,t_0)\Big)\\
\displaystyle + \Big( \theta_1 . V(x,t_0) + \theta_0 \Big)\times \mathds{1}_{[\frac{\theta_0}{1-\theta_1},+\infty[}\Big(V(x,t_0)\Big)
\end{array}
\tag{$\mathcal{M}_{2}$}
\end{equation}

We refer to model \ref{ltm} as the linear thresholded model since bad weather conditions have no consequence on road traffic flow under the critical speed $\theta_0/(1-\theta_1)$. Beyond this value, vehicles are decreasing their speeds linearly.\\

\subsection{An association between speeds}

\begin{figure}[!h]
  \centering
  \includegraphics[width=2.5in,angle = 90]{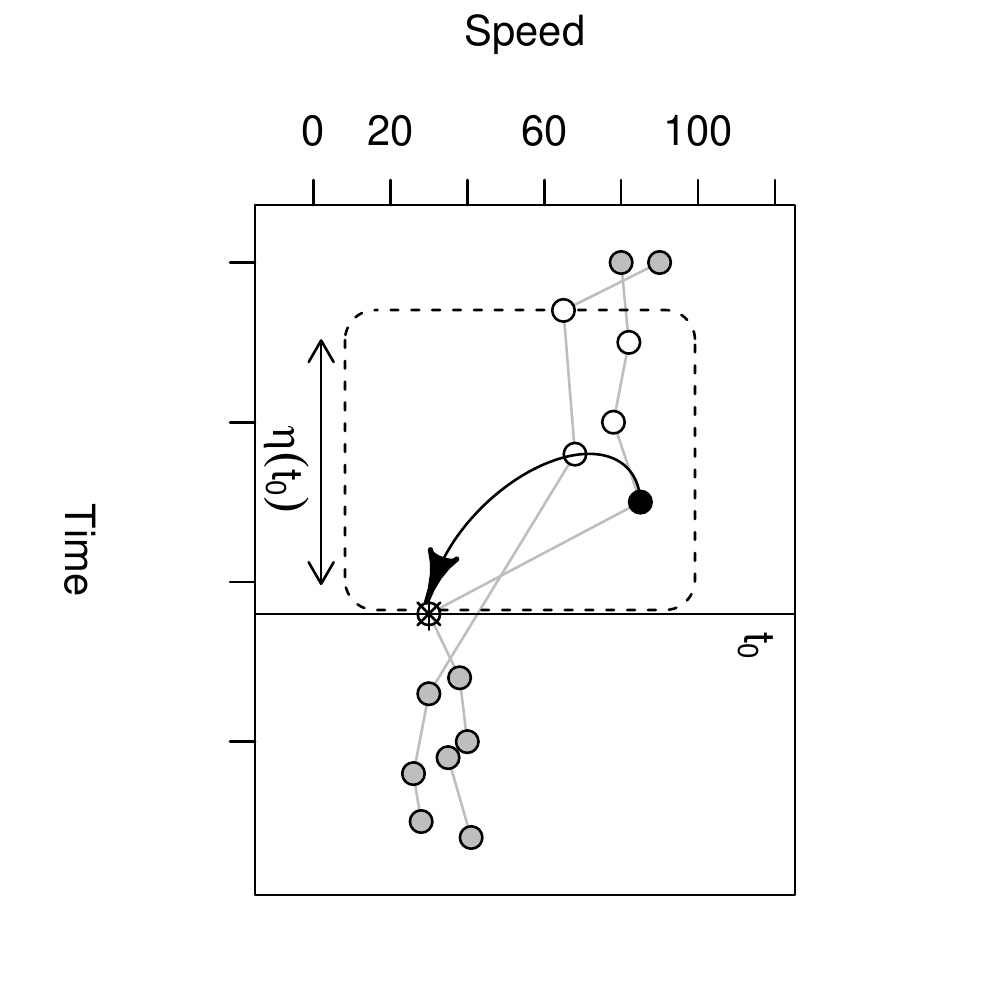}
  \caption{Matching between $V(x,t^*)$ and $V(x,t_0)$: the solid black line at $t_0$ corresponds to an occurrence of rain and $V(x,t_0)$, the ringed star, is the corresponding speed; white points correspond to points included in $\eta(t_0) = [t_0 - h, t_0]$ and $V(x,t^*)$, the black point, corresponds to the latest speed in $\eta(t_0)$}
  \label{correspBetweenSpeeds}
\end{figure}

We wish to build a matching between speeds of vehicles at a given location observed under the same traffic condition but with a different weather condition. So that, we use the following scheme (illustrated in Fig. \ref{correspBetweenSpeeds}):

\begin{enumerate}
\item extract occurrence times of climate change $t_0$,
\item build a time neighborhood $\eta(t_0)$ around $t_0$ in such a way that speeds in this neighborhood are stationary. In practice, we used to fix arbitrarily $\eta(t_0) = [t_0 - h, t_0]$ with $h = 5$ minutes. This is generally narrow enough to warrant stationarity,
\item let $t^*$ be the time of the latest observed speed before the climate change. Finding at least one observation in $\eta(t_0)$ for all $t_0$ is quite unlikely with our traffic data source. In fact, FCD are observed at random times so it is obvious that $t^*$ may not exist. The main consequence is a very sparse dataset. Thus, we decide to relax a little bit the assumption of traffic stationarity by allowing ourselves to pick similar velocities even if they belong to different observation days. This means that we assume the existence of a  stationarity between days. The only criterion that matters to pair the data is their belonging to the same temporal neighborhood $\eta(t_0)$. This ingenious consideration does not destroy more the stationarity than the climate change does itself,
\item associate $V(x,t^*)$ to $V(x,t_0)$ where $\mathcal{M}(x,t^*)\neq \mathcal{M}(x,t_0)$. For instance, if we want to study the impact of rain on microscopic speeds, $V(x,t^*)$ corresponds to speeds observed in no rain and no snow weather conditions (i.e. $\mathcal{M}(x,t^*) = $ NONE) and $V(x,t_0)$ correspond to speeds observed in the rain (i.e. $\mathcal{M}(x,t_0) = $ RAIN). 
\end{enumerate}

\section{Results} \label{s:three}
We have introduced in Section \ref{regression} two models that are able to represent the impact of bad weather conditions on microscopic speeds. In this section, we first aim to select the best link by link  model using a statistical approach. This gives a solution to a local modeling of our problem. Nevertheless, we also need to find a way to turn our link by link models into a network-wide model. This task is done right after the residual study for model selection.\\

The following results have been established with data presented in Section \ref{s:one} matching the region of Toulouse in France during 107 days from November $1^{st}$ 2009 to February $15^{th}$ 2010. We also focus on the impact of the rain because it is the most common adverse weather condition in France. So we basically work with \mbox{256 832} observations on
2070 links.

\subsection{Error}
We highlighted two models: the MARS model \ref{MARS} and the linear thresholded model \ref{ltm}. Here, we are facing a classical problem in statistics. We have to choose only one of these two models. What is the best model? How can we compare models in order to select the best one? Such questions are classical in model selection area and a solution can be set up.\\
We sample our data in two parts: a learning sample containing 90\% of observations to estimate our models and a test sample containing the remaining 10\% to validate and select the right model.\\

For each link of the network, we estimate both models by minimizing the Root Mean Squared Error (RMSE). The quality of a model is established by a classical goodness of fit value over all links calculated with the test sample

\begin{equation*}
\displaystyle RMSE = \sum_{x\in\{links\}} \sqrt{\frac{1}{n_x}\sum_{i=1}^{n_x}(\widehat{V(x,t)_i} - V(x,t)_i)^2}
\end{equation*}

\begin{figure}[!h]
  \centering
  \includegraphics[width=2.5in]{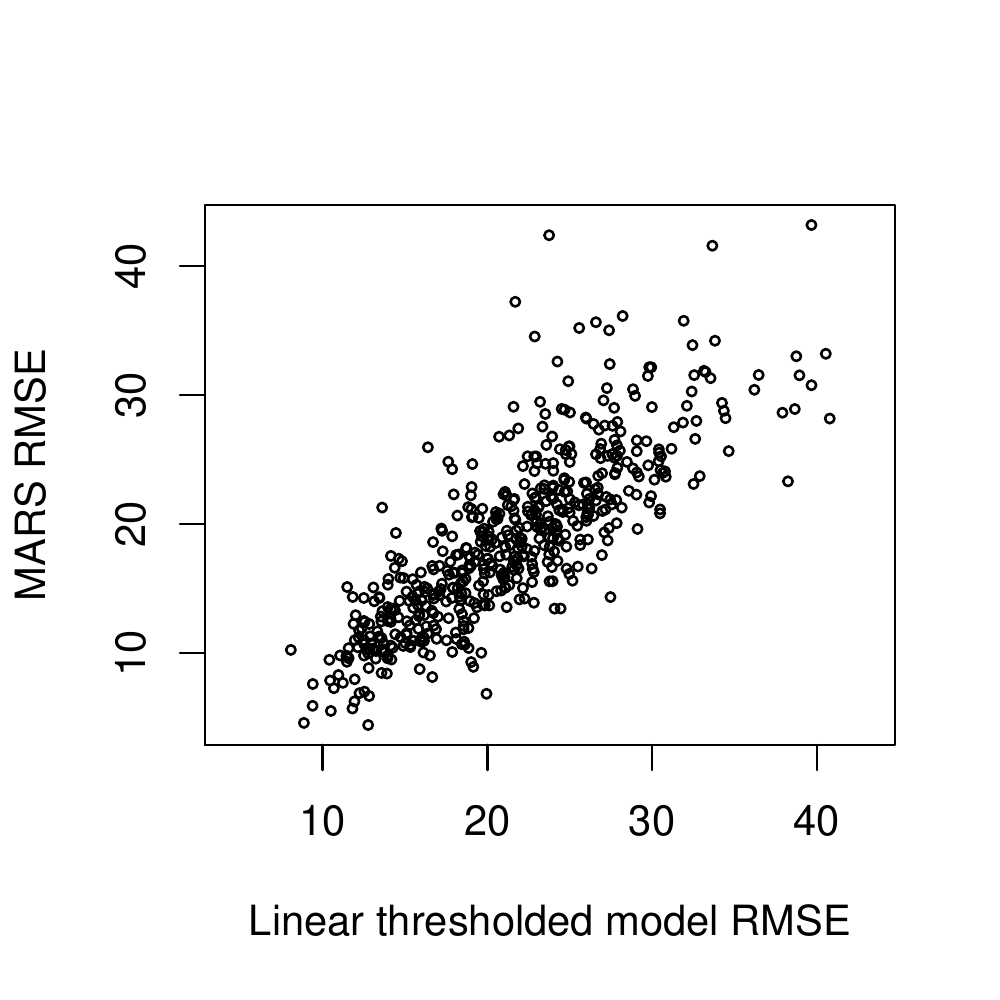}
  \caption{Biplot of RMSE by links for MARS model and linear thresholded model}
  \label{biplotResiduals}
\end{figure}

We obtain a RMSE of $10\,359.8$ for the MARS model and $11\,795.27$ for the linear thresholded model. So the MARS model fits the data better than the linear thresholded one but the difference of 1435.47 (13.8\%) is not significant since it has been calculated on 552 links. Moreover, the MARS model has a more complex structure than the linear thresholded one because the second is nothing more than a particular constrained MARS model. So it fits the data better by construction. Nevertheless, RMSE's on each link are similar for both models as shown in Fig. \ref{biplotResiduals}. This means that although we conclude that MARS model is better, the linear thresholded model is not so far behind and have the undeniable advantage of being extrapolatable to all a network whereas MARS models cannot because they have not a homogeneous structure among the network. This will be detailed in the next section. 

\subsection{Stability and extrapolation}
\label{stability}

We aim to set up a simple method to apply a correction to microscopic speeds based on adverse weather conditions on all a complex network at a time rather than on each link of a network taken separately. In fact, the France road network from FRC 0 to 3 is composed of \mbox{1 740 462} links and obviously there is not the same number of different behaviors in response of adverse weather conditions. Thus, our will of finding a global method is justified. In this section, we discuss about how we can generalized link dependent models to a global network model. Interests of extrapolating our models match our industrial constraints:
\begin{itemize}
\item to sum up and simplify our link by link models,
\item to apply a correction on speeds under adverse weather conditions on uncovered links.
\end{itemize} 

\begin{figure}[!h]
  \centering
  \includegraphics[width=2.5in]{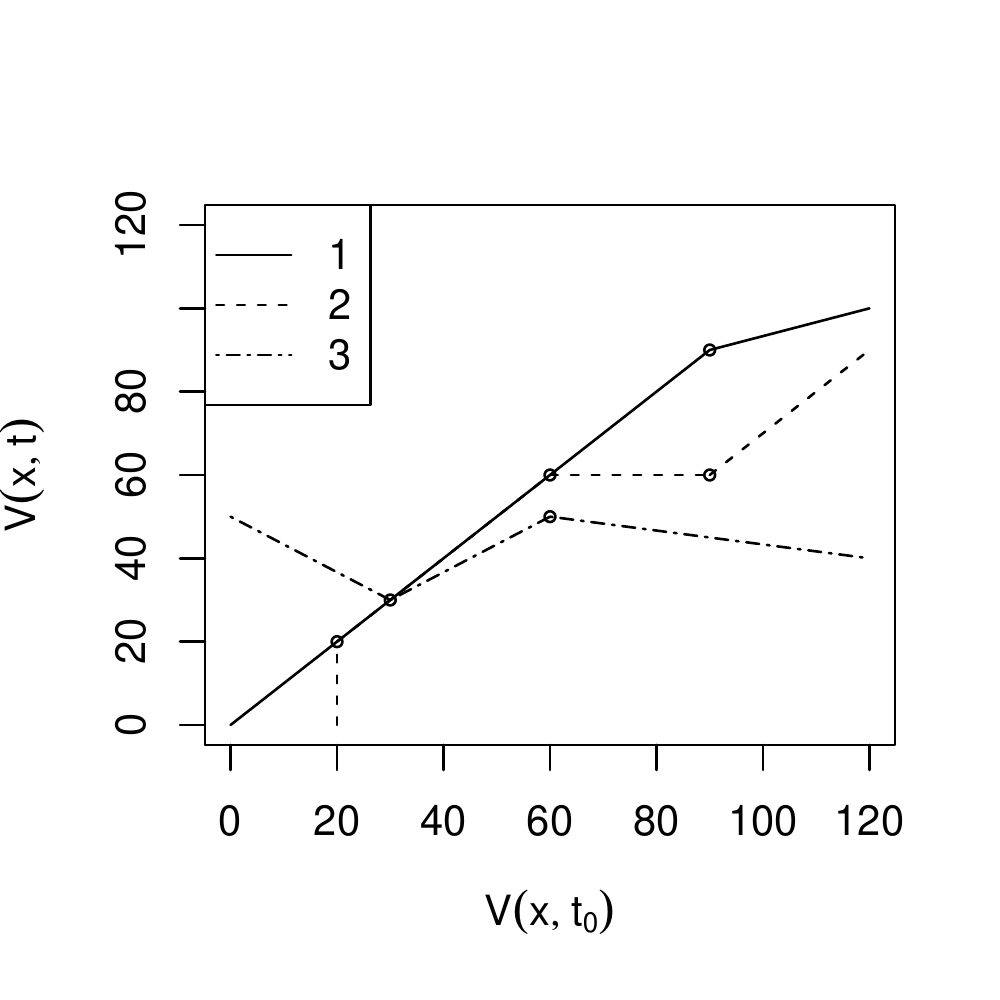}
  \caption{Example of three MARS models on three links}
  \label{ExampleofthreeMARSmodelsonthreeedges}
\end{figure}

First, we focus on the MARS model. For each link of the network, we have estimated a MARS model. We quickly face a problem to extrapolate this kind of model. Although this kind of model fits well the data, the structure is complex and different  among links.\\
For instance, Fig. \ref{ExampleofthreeMARSmodelsonthreeedges} shows three theoretical MARS models on three links. We observe that the number of slopes and the number of points to model non-linearities may differ from one link to another, making a global structure for a network difficult to build. Moreover, this kind of model is not consistent with road trafficking theory. As a matter of fact, under adverse road weather conditions drivers may reduce their velocities for high level speeds while their behavior is unaffected for low ones which cannot be always the case in general.\\

To sum up, link by link MARS models cannot be extrapolated to all a network mostly because they have not an homogeneous structure by link.\\

\begin{figure}[!h]
  \centering
  \includegraphics[width=2.5in]{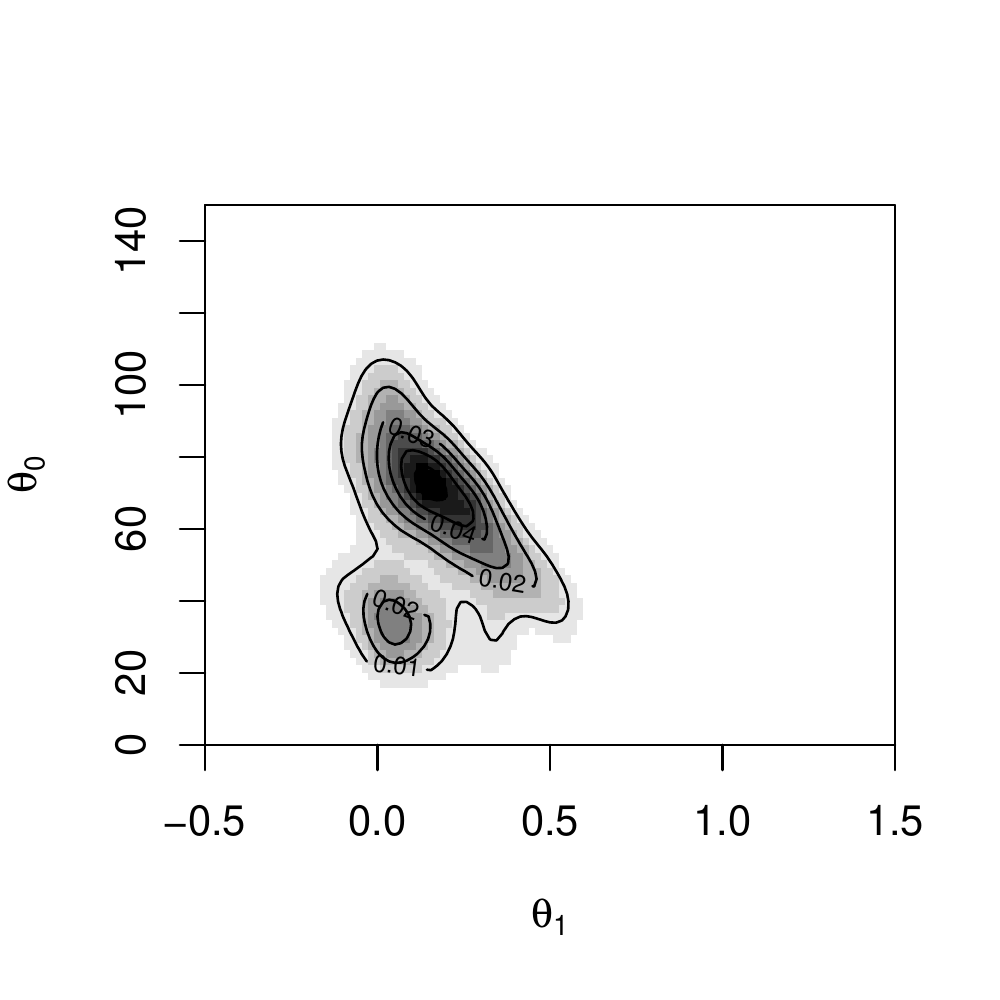}
  \caption{Global distribution of $(\theta_0,\theta_1)$}
  \label{Distributionofparametersforallvectices}
\end{figure}

\begin{figure}[!h]
  \centering
  \includegraphics[width=2.5in]{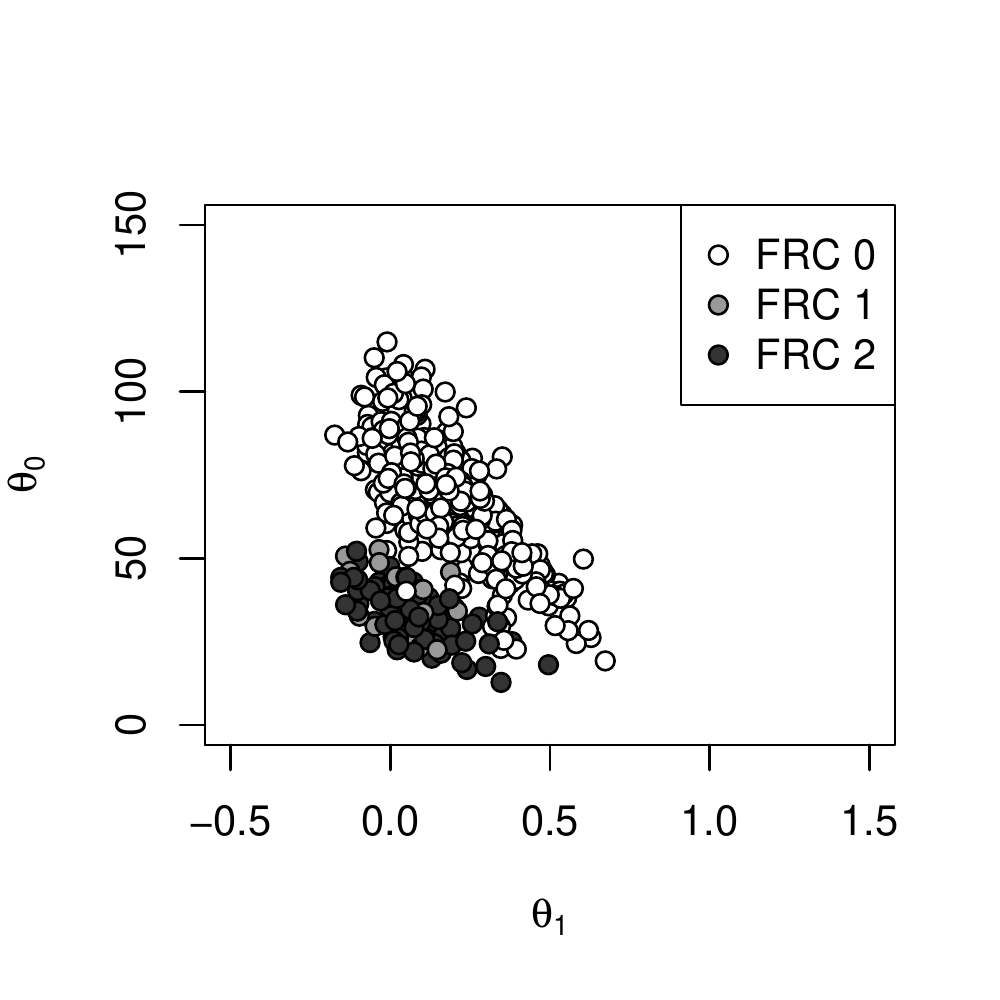}
  \caption{Biplot of $(\theta_0,\theta_1)$ for all links with their FRC}
  \label{DistributionofparametersforallvecticeswiththeirFRC}
\end{figure}

Second, we deal with linear thresholded models. Until now, our linear thresholded models were local because we built one model per link. With this kind of model, we always get a pair of parameters for each link which means that the structure is homogeneous among the network. So it is possible to extrapolate link by link models. Fig. \ref{Distributionofparametersforallvectices} shows a 2-dimensional kernel density estimation of the distribution of the two parameters for all these models. Two modes appear: one is associated to low $\theta_0$'s ( $\theta_0\leq 55$ km/h) and another to high ones ($\theta_0>55$ km/h). In fact, Fig. \ref{DistributionofparametersforallvecticeswiththeirFRC} shows exactly that marginal distribution for $\theta_0$ is highly related to the FRC.\\

\begin{figure}[!h]
  \centering
  \includegraphics[width=2.5in , angle = 90]{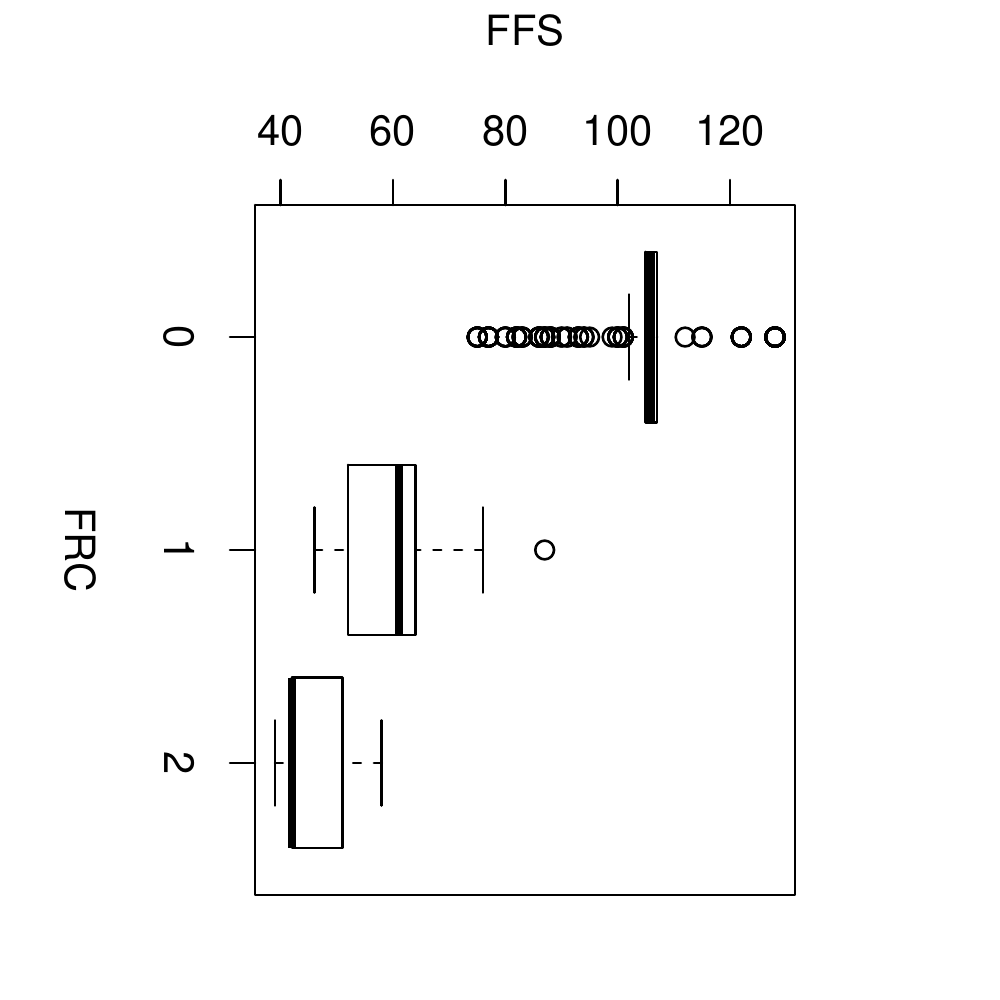}
  \caption{Boxplots of Free Flow Speeds by FRC}
  \label{BoxplotsoffreeflowspeedsbyFRC}
\end{figure}

We remind that we wish to build a global modeling for a network in other words our goal is to construct a model able to be applied on all the network. That is why we need to catch this dependence between $\theta_0$ and the FRC. A way to do that is to find a normalization $u$ such that the marginal distribution for $\tilde{\theta_0} = u(\theta_0,x)$ does not depend of the FRC. The use of $u(\cdot,x) = \cdot / \textrm{FFS}(x)$ will normalize $\theta_0$ correctly. Indeed, FFS is highly correlated with the FRC (see Fig. \ref{BoxplotsoffreeflowspeedsbyFRC}) and this normalization appears to be the most relevant and significant (5\% significant F-tests has been done).\\

\begin{figure}[!h]
  \centering
  \includegraphics[width=2.5in]{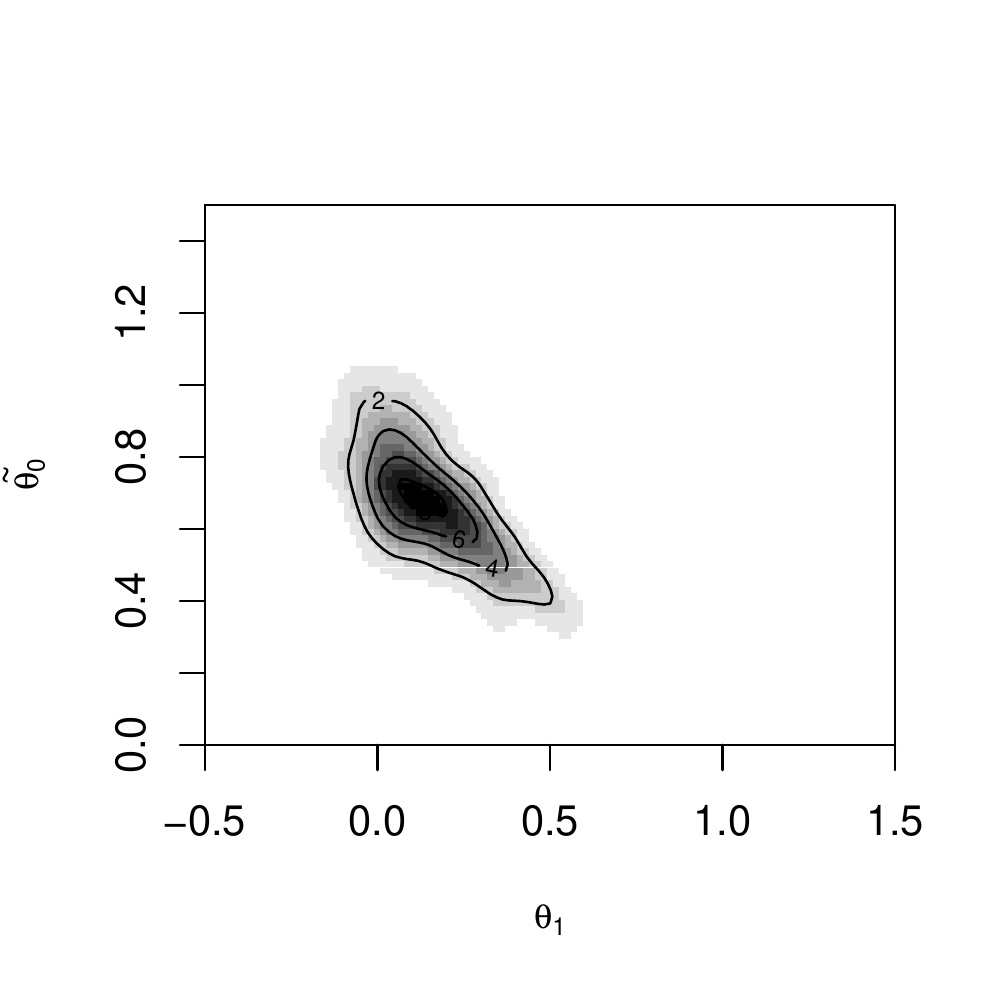}
  \caption{Global distribution of $(\theta_0,\tilde{\theta_1})$}
  \label{Distributionofnormalizedparametersforallvectices}
\end{figure}

Fig. \ref{Distributionofnormalizedparametersforallvectices} points out that the joint distribution $(\tilde{\theta_0},\theta_1)$ has been concentrated around one mode. We have thus stabilized the parameters over all FRC and thus over all the network.\\

Parameters of the global model remain as the empirical mean of the parameters of the link by link models i.e. $(\widehat{\theta_0},\widehat{\theta_1}) = (\overline{\tilde{\theta_0}}, \overline{\theta_1})$. One could naturally expect that $(\theta_0,\theta_1)$ is more than FRC dependent but also depend on the climate zone. Comparing the RMSE of the global model on climate  zones with the RMSE of models aggregated by climate zone, we concluded that the global  model was better in each case which means that the discrimination by climate zone is not significant.\\

Finally, our two link by link models \ref{MARS} and \ref{ltm} have similar RMSE on our data. Nevertheless, the MARS model can not be extrapolated by construction and it does not respect road trafficking theory. Thu,s the correct model for our problem is the linear thresholded model \ref{ltm} because it fits the data as the MARS model but it also respects road trafficking theory and it can be easily  extrapolated to the whole network. Constraining the MARS model makes it homogeneous among all the links and consequently extrapolable. The form of the global linear thresholded model for all the network remains

\begin{equation}
\label{ltmGlobal}
\begin{array}{l}
\displaystyle \forall t\in \mathcal{V}(t_0),\hspace{2em} V(x,t)  = \\ 
\displaystyle V(x,t_0) \times \mathds{1}_{[0,\frac{\overline{\tilde{\theta_0}}.FFS(x)}{1-\overline{\theta_1}}[}\Big(V(x,t_0)\Big)\\
\displaystyle + \Big( \overline{\theta_1} . V(x,t_0) + \overline{\tilde{\theta_0}}.FFS(x) \Big)\times
\mathds{1}_{[\frac{\overline{\tilde{\theta_0}}.FFS(x)}{1-\overline{\theta_1}},+\infty[}\Big(V(x,t_0)\Big)
\end{array}
\tag{$\mathcal{M}_{3}$}
\end{equation}

It is interesting to rewrite the model \ref{ltmGlobal} in a interpretable form for road trafficking experts. Let $\alpha = \frac{\overline{\tilde{\theta_0}}}{1- \overline{\theta_1}}$ and $\beta = 1 - \overline{\theta_1}$, we obtain\\

If $V(x,t_0) \geq \alpha.FFS(x)$,

\begin{equation}
\label{ltminterpretableForm}
\begin{array}{c}
V (x,t) = V(x,t_0) - \beta.\Big(V(x,t_0) - \alpha.FFS(x) \Big)\\
\end{array}
\tag{$\mathcal{M}_{4}$}
\end{equation}

Else
\begin{equation*}
\begin{array}{c}
V(x,t) = V(x,t_0)
\end{array}
\end{equation*}

\begin{figure}[!h]
  \centering
  \includegraphics[width=2.5in]{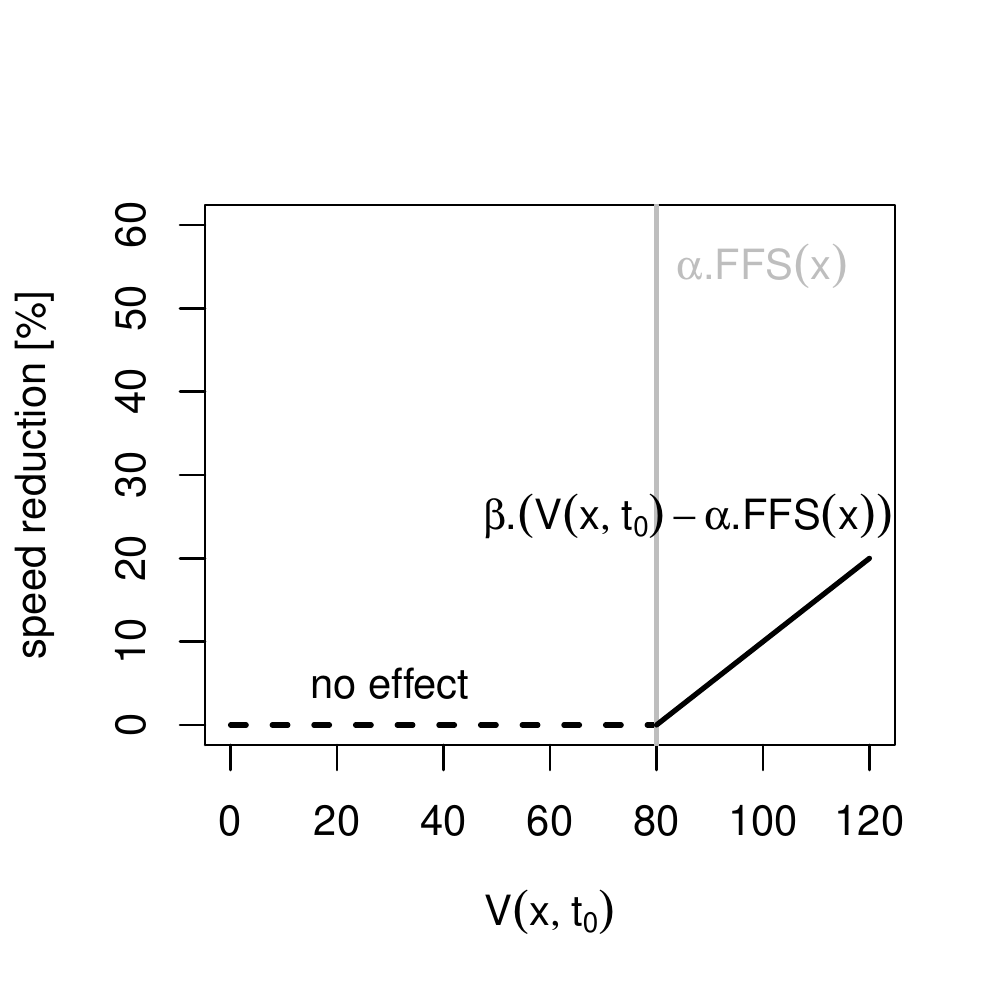}
  \caption{Percentage of speed reduction based on \ref{ltminterpretableForm}}
  \label{explanationCorrection}
\end{figure}

Thus the interpretation is really natural and illustrated in Fig. \ref{explanationCorrection}. If there is a vehicle at a speed above a proportion of the Free Flow Speed $\alpha.FFS(x)$ and it starts raining, we decrease its speed by $\beta.\Big(V(x,t_0) - \alpha.FFS(x) \Big)$ i.e. a proportion of the difference between its initial speed and the proportion of the Free Flow Speed . $\alpha.FFS(x)$ represents the speed at which adverse weather conditions start impacting speeds.\\

We have built a global model which adapts itself locally since we use this ingenious normalization $\tilde{\theta_0} = \theta_0/FFS(x)$. With our data, estimates of \ref{ltmGlobal} are $(\overline{\tilde{\theta_0}} , \overline{\theta_1}) = (0.66,0.16)$. Let us consider a basic example to practice with the model: a vehicle is recorded at $130$ km/h on a freeway where the Free Flow Speed is $130$ km/h and it starts raining. So since $130 \geq 0.66\times 130/(1-0.16)$, we apply a correction and the speed is reduced to $130 - (016\times 130 + 0.66\times 130) = 106.6$ km/h. Remark that it respects the French road speed limit on a freeway ($130$ km/h in general and decreasing to $110$ km/h when raining).\\

Until now, we have selected the linear thresholded model and extrapolated it on a network. We know that our extrapolation by normalizing $\theta_0$ by the Free Flow Speed was the best relevant according to our result but we also need to measure the actual loss of quality in extrapolating the model. So we calculate and compare these two following quantities still calculated on the test sample

\begin{equation*}
\label{qualityExtrapolation}
\begin{array}{c}
\displaystyle RMSE_{|\textrm{\ref{ltm}}} = \sum_{x\in\{links\}} \sqrt{\frac{1}{n_x}\sum_{i=1}^{n_x}(\widehat{V(x,t)_i}_{|\textrm{\ref{ltm}}} - V(x,t)_i)^2}\\
\displaystyle RMSE_{|\textrm{\ref{ltmGlobal}}} = \sum_{x\in\{links\}}\sqrt{\frac{1}{n_x}\sum_{i=1}^{n_x}(\widehat{V(x,t)_i}_{|\textrm{\ref{ltmGlobal}}} - V(x,t)_i)^2}
\end{array}
\end{equation*}

with $\widehat{V(x,t)_i}_{|\textrm{\ref{ltm}}}$ the forecasted speed with a link by link linear thresholded model and $\widehat{V(x,t)_i}_{|\textrm{\ref{ltmGlobal}}}$ the forecasted speed with a linear thresholded model for the network i.e. link by link models extrapolated to all the network.\\

We obtain $\displaystyle RMSE_{|\textrm{\ref{ltm}}} = 11\,795.27$ and $\displaystyle RMSE_{|\textrm{\ref{ltmGlobal}}} = 12\,511.37$. The loss of quality associated to the globalization $\displaystyle \Delta RMSE$ is equal to $6.07 \%$. Thus we can warrant that the extrapolation of our link by link models is relevant and do not destroy the quality of fit compared to our local models.\\

\section{Conclusion}\label{s:conclusion}

We have tackled the issue of building a generic rule able to predict the evolution of vehicle velocities when the weather condition changes. For this, we have considered a version of Multivariate Adaptive Regression Spline model, well calibrated to get stable and accurate predictions. So we get model \ref{ltminterpretableForm} that basically does not modify speeds under a proportion of the free flow speed. Above this threshold value, the model decreases the speed by a proportion of the difference between this speed and the threshold value. Thus, one can use the model \ref{ltminterpretableForm} to correct forecasted speeds with the information of weather conditions. To learn the model over a data set, we had to construct a well adapted learning set. One of the difficulties of this task was to overcome the non stationarity of the observed data which mix both the variability due to the changes of the weather conditions and the one due to the changes in the traffic conditions.  This was achieved by considering time neighborhood around similar velocities.  Moreover the desired stability of the  decision rule enables us to extend this method over a whole road network. This is, to our knowledge, the first global quantitative analysis of the impact of adverse weather on the observed vehicle velocities. This is a major improvement to forecast travel time with some knowledge on the weather conditions. This contributes to a better quality of the forecasts done by Mediamobile company.\\

Moreover, this study is also a key to better understand the macroscopic impact of adverse weather conditions on the Free Flow Speed. Actually,  some fancy results emerge with the extrapolation of the linear thresholded model. We could think that adverse weather conditions not only impact microscopic speeds but also impact the well known Free Flow Speed. The global model \ref{ltmGlobal} includes such a result. When building this model, we build a speed at which adverse weather conditions start impacting speeds and this speed is nothing more than a proportion of the Free Flow Speed $\alpha.FFS(x)$. In this work, we provide another reference speed which can be considered as the Free Flow Speed under a certain adverse weather condition.

\section*{Acknowledgment}

The authors would like to thank M\'et\'eo-France and \mbox{Mediamobile} for providing respectively weather and road traffic data through their Road\&Weather partnership.

\ifCLASSOPTIONcaptionsoff
  \newpage
\fi

\bibliographystyle{IEEEtran}

\bibliography{./ModelingWeatherConditionsConsequencesOnRoadTraffickingBehaviors}

\begin{thebibliography}{10}
\providecommand{\url}[1]{#1}
\csname url@samestyle\endcsname
\providecommand{\newblock}{\relax}
\providecommand{\bibinfo}[2]{#2}
\providecommand{\BIBentrySTDinterwordspacing}{\spaceskip=0pt\relax}
\providecommand{\BIBentryALTinterwordstretchfactor}{4}
\providecommand{\BIBentryALTinterwordspacing}{\spaceskip=\fontdimen2\font plus
\BIBentryALTinterwordstretchfactor\fontdimen3\font minus
  \fontdimen4\font\relax}
\providecommand{\BIBforeignlanguage}[2]{{%
\expandafter\ifx\csname l@#1\endcsname\relax
\typeout{** WARNING: IEEEtran.bst: No hyphenation pattern has been}%
\typeout{** loaded for the language `#1'. Using the pattern for}%
\typeout{** the default language instead.}%
\else
\language=\csname l@#1\endcsname
\fi
#2}}
\providecommand{\BIBdecl}{\relax}
\BIBdecl

\bibitem{Friedman1993}
\BIBentryALTinterwordspacing
J.~H. Friedman, ``{Fast MARS},'' Department of Statistics and Stanford Linear
  Accelerator Center - Stanford University, Stanford, Tech. Rep., 1993.
  [Online]. Available:
  \url{http://statistics.stanford.edu/~ckirby/techreports/LCS/LCS 110.pdf}
\BIBentrySTDinterwordspacing

\bibitem{ElFaouzi2008}
N.-E. {El Faouzi}, O.~de~Mouzon, and R.~Billot, ``{Toward Weather-Responsive
  Traffic Management on French Motorways},'' \emph{Transportation Research
  Circular}, pp. 443--456, 2008.

\bibitem{Elfaouzi2010}
R.~B.~S. {El Faouzi N. -E. Billot}, ``{Motorway travel time prediction based on
  toll data and weather effect integration},'' \emph{IET INTELLIGENT TRANSPORT
  SYSTEMS}, vol.~4, pp. 338--345, 2010.

\bibitem{Elfaouzi2010Bis}
N.-E. {El Faouzi}, R.~Billot, P.~Nurmi, and B.~Nowotny, ``{Effects of adverse
  weather on traffic and safety: State-of-the-art and a European initiative},''
  in \emph{15th International Road Weather Conference}, 2010.

\bibitem{Billot2009}
\BIBentryALTinterwordspacing
R.~Billot, N.-E. {El Faouzi}, and F.~{De Vuyst}, ``{Multilevel Assessment of
  the Impact of Rain on Drivers' Behavior},'' \emph{Transportation Research
  Record}, vol. 2107, no.~-1, pp. 134--142, 2009. [Online]. Available:
  \url{http://trb.metapress.com/openurl.asp?genre=article\&id=doi:10.3141/2107-14}
\BIBentrySTDinterwordspacing

\bibitem{Billot2010}
R.~Billot and J.~Sau, ``{Integrating the impact of rain into traffic management
  : online traffic state estimation using sequential Monte Carlo techniques},''
  \emph{Transportation Research Record}, pp. 1--14, 2010.

\bibitem{Kilpelainen2007}
\BIBentryALTinterwordspacing
M.~Kilpel\"{a}inen and H.~Summala, ``{Effects of weather and weather forecasts
  on driver behaviour},'' \emph{Transportation Research Part F Traffic
  Psychology and Behaviour}, vol.~10, no.~4, pp. 288--299, 2007. [Online].
  Available:
  \url{http://linkinghub.elsevier.com/retrieve/pii/S1369847806000982}
\BIBentrySTDinterwordspacing

\bibitem{Friedman1991}
\BIBentryALTinterwordspacing
J.~H. Friedman, ``{Multivariate adaptive regression splines},'' \emph{Ann.
  Statist.}, vol.~19, no.~1, pp. 1--141, 1991. [Online]. Available:
  \url{http://dx.doi.org/10.1214/aos/1176347963}
\BIBentrySTDinterwordspacing

\bibitem{Garthwaite1997}
\BIBentryALTinterwordspacing
P.~H. Garthwaite, P.~J. Brown, D.~J. Hand, S.~Wold, D.~R. Cox, J.~V. Zidek,
  C.~J.~F. TerBraak, M.~Stone, R.~Brooks, C.~Goutis, D.~V. Lindley, A.~J.
  Burnham, J.~F. MacGregor, R.~Viveros, T.~Hastie, R.~Tibshirani, I.~S.
  Helland, M.~C. Jones, P.~D. Sasieni, R.~Southworth, C.~C. Taylor,
  R.~Sundberg, E.~V. Thomas, and H.~Tong, ``{Predicting multivariate responses
  in multiple linear regression - Discussion},'' \emph{Journal of the Royal
  Statistical Society - Series B: Statistical Methodology}, vol.~59, no.~1, pp.
  3--54, 1997. [Online]. Available: \url{http://kar.kent.ac.uk/18067/}
\BIBentrySTDinterwordspacing

\bibitem{Mammen1997}
E.~Mammen and S.~{Van De Geer}, ``{Locally Adaptive Regression Splines},''
  \emph{Annals of Statistics}, vol.~25, pp. 387--413, 1997.

\bibitem{Poovadan2011}
A.~K. Poovadan, \emph{{MultiNet EUR 2011.12 Product Release Notes}}, T.~N.
  America, Ed.\hskip 1em plus 0.5em minus 0.4em\relax TomTom North America,
  2011.

\bibitem{Loubes2006}
\BIBentryALTinterwordspacing
J.-M. Loubes, E.~Maza, M.~Lavielle, and L.~Rodriguez, ``{Road trafficking
  description and short term travel time forecasting, with a classification
  method},'' \emph{Canadian Journal Of Statistics}, vol.~34, no.~3, pp.
  475--491, 2006. [Online]. Available:
  \url{http://doi.wiley.com/10.1002/cjs.5550340307}
\BIBentrySTDinterwordspacing

\bibitem{Allain2009}
G.~Allain, F.~Gamboa, P.~Goudal, J.-M. Loubes, and E.~Maza, ``{A Statistical
  Framework for Road Traffic Prediction},'' in \emph{16th ITS World Congress
  and Exhibition on Intelligent Transport Systems and Services}, 2009.

\end{thebibliography}









\end{document}